\documentclass{INTERSPEECH2023}

\interspeechcameraready

\usepackage{amsmath,graphicx,enumitem,amsfonts,multirow,tabularx,makecell,bbm,algorithm,algpseudocode,hyperref,url,booktabs,amssymb,pifont,arydshln}
\usepackage[svgnames]{xcolor}
\usepackage[labelfont=bf, labelsep=colon]{caption}

\DeclareMathOperator*{\argmax}{\arg\max}

\def\y{{\boldsymbol y}}

\newcommand{\ie}{\textit{i}.\textit{e}.}

\definecolor{applegreen}{rgb}{0.55, 0.71, 0.0}
\definecolor{custombrown}{rgb}{0.79, 0.64, 0.45}
\definecolor{lightgray}{rgb}{0.7, 0.7, 0.7}
\newcommand{\cmark}{\ding{51}}

\newcommand*\colourcheck[1]{
  \expandafter\newcommand\csname #1check\endcsname{\textcolor{#1}{\ding{51}}}%
}
\newcommand*\colourx[1]{
  \expandafter\newcommand\csname #1x\endcsname{\textcolor{#1}{\ding{55}}}%
}

\newcommand{\blank}{{\mspace{1mu}\cdot\mspace{1mu}}}

\algrenewcommand\algorithmicrequire{\textbf{Input:}}
\algrenewcommand\algorithmicensure{\textbf{Output:}}
\algnewcommand\Input{\item[\algorithmicinput]}
\algnewcommand\Output{\it em[\algorithmicoutput]}

\hypersetup{colorlinks,urlcolor=blank}
\colourx{lightgray}

\title{SGEM: Test-Time Adaptation for Automatic Speech Recognition via Sequential-Level Generalized Entropy Minimization}
\name{Changhun Kim$^1$ \quad Joonhyung Park$^1$ \quad Hajin Shim$^1$ \quad Eunho Yang$^{1,2}$}
\address{
  $^1$Korea Advanced Institute of Science and Technology (KAIST), South Korea \\
  $^2$AITRICS, South Korea
}
\email{\{changhun.kim, deepjoon, shimazing, eunhoy\}@kaist.ac.kr}

\begin{document}

\maketitle
 
\begin{abstract}
Automatic speech recognition (ASR) models are frequently exposed to data distribution shifts in many real-world scenarios, leading to erroneous predictions. To tackle this issue, an existing test-time adaptation (TTA) method has recently been proposed to adapt the pre-trained ASR model on unlabeled test instances without source data. Despite decent performance gain, this work relies solely on naive greedy decoding and performs adaptation across timesteps at a frame level, which may not be optimal given the sequential nature of the model output. Motivated by this, we propose a novel TTA framework, dubbed SGEM, for general ASR models. To treat the sequential output, SGEM first exploits beam search to explore candidate output logits and selects the most plausible one. Then, it utilizes generalized entropy minimization and negative sampling as unsupervised objectives to adapt the model. SGEM achieves state-of-the-art performance for three mainstream ASR models under various domain shifts.\footnote{\scriptsize{Our code is available at \url{https://github.com/drumpt/SGEM}.}}
\end{abstract}

\section{Introduction}
\label{sec:intro}
While deep neural networks have achieved remarkable progress in a broad range of areas, such as computer vision~\cite{he2016deep, vit}, natural language processing~\cite{vaswani2017attention, gpt3}, and speech processing~\cite{shen2018natural, baevski2020wav2vec}, these models are known to be susceptible to data distribution shifts~\cite{adda, uda_back}. This so-called \textit{domain shift problem} readily occurs in recent automatic speech recognition (ASR) models, which imposes challenges in deploying ASR models to real-world applications. For example, utterances of unseen speakers/words not exposed during training or utterances with accidental background noise can be given at test time.

To tackle this domain shift problem for ASR models, many prior works have been suggested, including data augmentation~\cite{hsu2017unsupervised}, feature alignment~\cite{hou2021cross}, domain adversarial learning~\cite{sun2017unsupervised, sun2018domain}, and knowledge distillation~\cite{manohar2018teacher}. These works mainly mitigate the domain shift problem under the unsupervised domain adaptation (UDA) setting, where the source models are adapted to unlabeled target domains. However, the UDA setting has several impractical assumptions in real-world scenarios. First, it assumes that the source data is accessible, which might be unavailable due to privacy/storage issues. Second, a pile of target data has to be collected \textit{in advance}. This is also unrealistic as it requires substantial resources. Even worse, it restricts the generalization capacity of the model only to the pre-collected target data, although the target distribution can change arbitrarily at test time. Meanwhile, several speaker adaptation methods~\cite{seide2011feature, yao2012adaptation, yu2013kl} have demonstrated satisfactory adaptation performance on variation in speakers during inference. However, their restricted focus on speaker changes and reliance on prior knowledge of test-time speakers impose limitations in effectively addressing arbitrary domain shifts during test time.

Inspired by these limitations, a method called SUTA~\cite{suta} was first proposed to address an arbitrary domain shift problem under a more realistic setting - test-time adaptation (TTA) - for ASR models as in other domains~\cite{sun2020test, wang2020tent, liu2021ttt++, fleuret2021test, zhang2021memo, chen2022contrastive, wang2022continual, bartler2022mt3, goyal2022test, niu2023towards}. Given an off-the-shelf ASR model pre-trained on the source domain, TTA methods aim to adapt the model on-the-fly using unlabeled instances from the target domain in test time without access to source data. Inheriting the ideas of the TTA approaches in the computer vision domain~\cite{wang2020tent, fleuret2021test, zhang2021memo}, SUTA shows decent performance in a single-utterance TTA setting for the CTC-based ASR model~\cite{baevski2020wav2vec}.

However, directly adopting this approach to advanced ASR models~\cite{burchi2021efficient, radford2022robust} could not be optimal as SUTA was developed with the CTC-based models~\cite{graves2006connectionist} in mind. This is because, unlike the CTC-based models, which generate each output token independently in a greedy manner, advanced ASR models are designed to work in an autoregressive manner or typically utilize beam search decoding with an external language model during the test phase. This indicates that output logits acquired by greedy decoding may not adequately capture the output distribution, and naively adapting with these logits at a frame level as in~\cite{suta} can cause undesirable behavior at the sequential level for general ASR models.

In this paper, we propose a \textbf{S}equential-Level \textbf{G}eneralized \textbf{E}ntropy \textbf{M}inimization (SGEM) framework towards an effective TTA for general ASR models. To this end, SGEM first explores candidate output logits and selects the most plausible one using beam search to leverage the sequential nature of the output. Then, SGEM leverages generalized entropy minimization loss and negative sampling loss as auxiliary unsupervised objectives to adapt the model at a sequential level. We validate SGEM for three representative ASR models on various datasets with different distribution shifts, including unseen speakers/words and severe background noise, and demonstrate that SGEM achieves state-of-the-art performance in most settings. To the best of our knowledge, this is the first work suggesting the TTA method for general ASR models.

\begin{figure*}[!t]
    \centering
    \includegraphics[width=.9\textwidth]{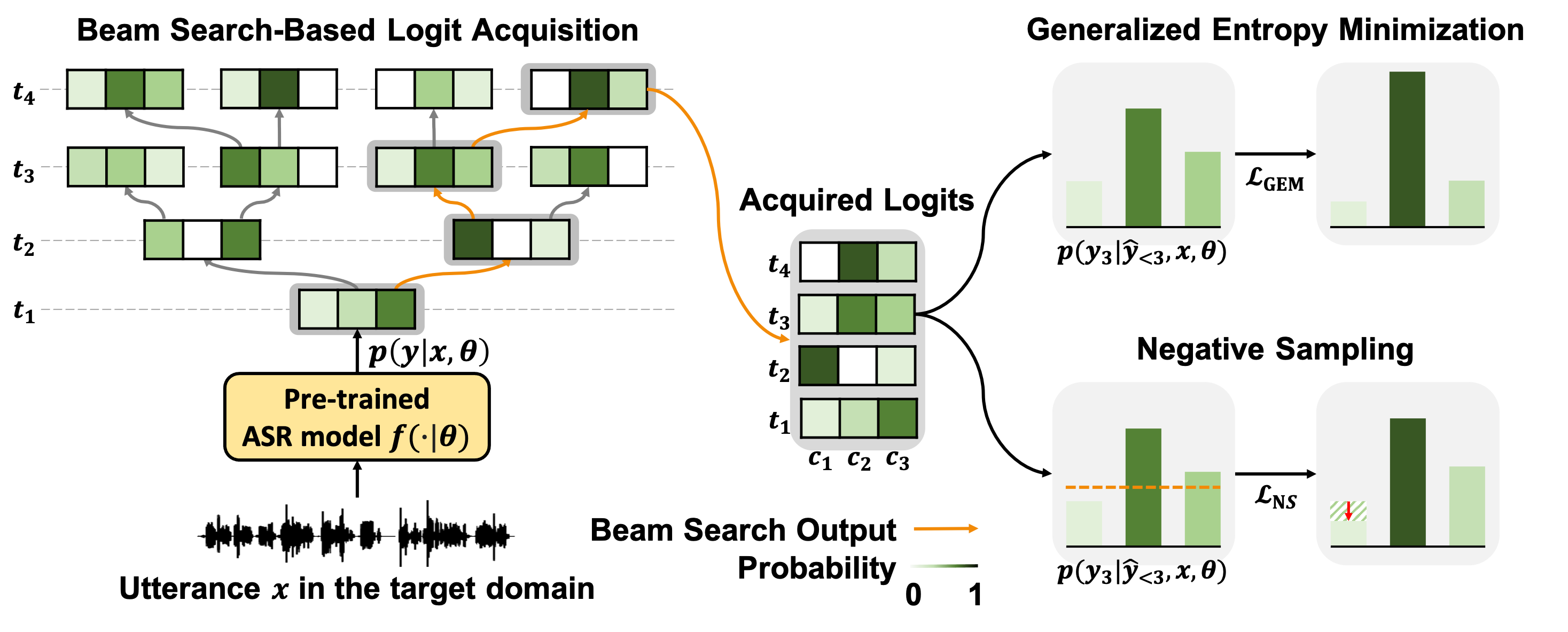}
    \vspace{-0.1in}
    \caption{The overall pipeline of the SGEM framework. Given a single utterance $x$ in the target domain, we first acquire logits based on the most plausible beam search output. Then, we utilize generalized entropy minimization and negative sampling objectives with $x$ itself to adapt the ASR model $f(\cdot|\theta)$ pre-trained on the source domain.}
    \label{fig:overview}
    \vspace{-0.25in}
\end{figure*}

\section{Proposed Method: SGEM}
\label{sec:method}
This section introduces SGEM, an effective TTA framework for general ASR models. To this end, we describe the test-time adaptation setup for ASR models in Section~\ref{subsec:tta_setup} and illustrate the core strategies of SGEM from Section~\ref{subsec:logit_acquisition} to Section~\ref{subsec:overall_framework}. Figure~\ref{fig:overview} depicts the overall pipeline of the proposed method.

\subsection{Test-Time Adaptation Setup for ASR Models}
\label{subsec:tta_setup}
We briefly explain general ASR models before formulating test-time adaptation (TTA). Let $f(\cdot | {\theta})$ be an ASR model trained on the labeled source domain $\mathcal{D}_s = {\{(x_i^{s}, y_i^{s} )\}}_{i}$ in pairs of speech and text, which takes a raw waveform $x$ and returns logits $f(x | \theta) \in \mathbb{R}^{L \times C}$ for each timestep. Here, $L$ is the number of timesteps, $C$ is the number of vocabulary classes, and $\theta$ is the model parameter. It models the log joint probability $\log p(\y | x, \theta)$ of a candidate transcript $\y = (y_1, \cdots, y_{L})$, in an \textit{autoregressive manner} as follows:
\begin{align*}
&\log p(\y | x, \theta) := \log p_{\text{AM}} (\y| x, \theta) + \lambda_{\text{LM}} \log p_{\text{LM}}(\y) + Z \\
&= \sum_{i=1}^{L} \log p_{\text{AM}} (y_i | y_{<i}, x, \theta) + \lambda_{\text{LM}} \log p_{\text{LM}}(y_i | y_{<i}) + Z,
\end{align*}
where $y_i \in \{1, \cdots, C\}$, $p_{\text{AM}}(\y| x, \theta)$ is the joint probability given by the model output $f(x | \theta)$, ~$p_{\text{LM}}(\y)$ is the joint probability of an autoregressive language model (LM), $\lambda_{\text{LM}}$ is a hyperparameter to control the effect of the LM, and $Z$ is a normalizing constant. LM aims to boost the ASR model to generate more faithful sentences. ASR decoding strategies approximate the optimal solution $\y^{*} = \argmax_{\y} \log p(\y | x, \theta)$.

TTA methods for an ASR model $f(\cdot | \theta)$ aim to adapt the model to the unlabeled target speech domain $\mathcal{D}_t = {\{ x_i^{t} \}}_{i}$ without access to $\mathcal{D}_s$. Specifically, we consider a \textit{single-utterance TTA setting} where we fine-tune the ASR model $f(\cdot|\theta)$ for each utterance $x_{i}^{t} \in \mathcal{D}_t$ to get more precise output logits ${\log p(\y | x_{i}^{t}, \theta)}$ with unsupervised objectives using only $x_i^t$ itself. This single-utterance TTA setting is considerably pragmatic regarding low latency without presuming that the test instances are independent and identically distributed~\cite{suta, wang2022continual, niu2023towards}.

\begin{table*}[!t]
\scriptsize
\centering
\caption{Comparison of TTA performance, measured as word error rate (\%), for three mainstream ASR models across 12 datasets with various types of domain shifts. The results are obtained using greedy decoding for inference.}
\label{table:main}
\vspace{-0.1in}
\begin{tabular}{clccccccccccccc}
\Xhline{3\arrayrulewidth}
    & \textbf{Dataset} & \textbf{CH} & \textbf{TD} & \textbf{CV} & \textbf{VA} & \textbf{AC} & \textbf{AA} & \textbf{BA} & \textbf{CM} & \textbf{MU} & \textbf{NB} & \textbf{SD} & \textbf{TP} & \textbf{Avg.} \\ \cline{1-15}
    \multirow{3}{*}{\begin{tabular}[c]{@{}c@{}}\textbf{CTC-based}\\\textbf{model}\end{tabular}} & Unadapted & 31.2 & 13.2 & 36.9 & 14.5 & 28.1 & 40.9 & 66.6 & 49.8 & 50.4 & 119.2 & 19.2 & 26.2 & 41.4 \\ \cline{2-15}
    & SUTA & 25.0 & \textbf{12.0} & 31.4 & 11.8 & 17.7 & 31.3 & 55.2 & 39.4 & 39.7 & 113.0 & 15.0 & 17.8 & 34.1 \\ \cline{2-15} 
    & SGEM & \textbf{24.7} & \textbf{12.0} & \textbf{31.1} & \textbf{11.6} & \textbf{17.3} & \textbf{30.7} & \textbf{53.1} & \textbf{38.5} & \textbf{38.6} & \textbf{110.5} & \textbf{14.8} & \textbf{17.5} & \textbf{33.4} \\ \cline{1-15} \noalign{\vskip\doublerulesep \vskip-\arrayrulewidth} \cline{1-15}
    \multirow{3}{*}{\textbf{Conformer}} & Unadapted & 28.7 & 15.1 & 36.8 & 17.4 & 18.8 & 44.8 & 74.3 & 45.7 & 56.0 & 122.1 & 20.8 & 36.9 & 43.1 \\ \cline{2-15}
    & SUTA & 25.2 & 13.4 & 32.4 & 14.7 & 14.5 & 39.8 & 73.3 & \textbf{38.4} & 48.7 & 125.5 & \textbf{16.4} & \textbf{28.8} & 39.3 \\ \cline{2-15}
    & SGEM & \textbf{24.5} & \textbf{13.3} & \textbf{31.6} & \textbf{14.6} & \textbf{14.4} & \textbf{38.5} & \textbf{70.4} & 38.7 & \textbf{48.5} & \textbf{120.9} & 16.8 & 28.9 & \textbf{38.4} \\ \cline{1-15} \noalign{\vskip\doublerulesep \vskip-\arrayrulewidth} \cline{1-15}
    \multirow{3}{*}{\textbf{Transducer}} & Unadapted & 11.8 & 7.2 & 12.9 & 6.5 & 14.1 & 20.4 & 31.0 & 29.7 & 31.3 & \textbf{74.6} & 12.7 & 16.2 & 22.4 \\ \cline{2-15}
    & SUTA & 10.3 & 6.8 & 12.1 & 5.5 & 12.0 & 18.5 & 28.3 & 26.7 & 28.7 & \textbf{74.6} & 11.7 & 14.7 & 20.8 \\ \cline{2-15} 
    & SGEM & \textbf{9.9} & \textbf{6.6} & \textbf{12.0} & \textbf{5.2} & \textbf{11.6} & \textbf{18.0} & \textbf{27.5} & \textbf{26.0} & \textbf{28.0} & 76.5 & \textbf{11.5} & \textbf{14.3} & \textbf{20.6} \\
\Xhline{3\arrayrulewidth}
\vspace{-0.3in}
\end{tabular}
\end{table*}

\subsection{Beam Search-Based Logit Acquisition} 
\label{subsec:logit_acquisition}
An existing TTA method for ASR models~\cite{suta} exploits the greedy decoding strategy without leveraging the external LM (\ie, $\lambda_{\text{LM}} = 0$) to get output logits for all timesteps. Also, it utilizes them with unsupervised objectives such as entropy minimization~\cite{grandvalet2004semi} and minimum class confusion~\cite{jin2020minimum} as if they are appropriate logits to adapt the ASR model. However, naively using greedy decoding is proven defective~\cite{freitag2017beam} and can mislead the model to be adapted on the wrong labels. Furthermore, this frame-level greedy adaptation might be sub-optimal on the sequential output since it only considers the joint probability of a sequence myopically over timesteps.

To this end, we exploit a novel logit acquisition strategy based on more systematic beam search decoding. Given a beam width $B$, we find the most plausible output sequence $\hat{\y} = (\hat{y}_{1}, \cdots, \hat{y}_{L})$, which approximates $\y^{*}$ using beam search~\cite{freitag2017beam}. Note that we do not hold the logits of beam candidates in this step to reduce memory consumption. Instead, the estimated sequence $\hat{\y}$ is passed to the model again to acquire the $i$-th logit $\boldsymbol{o}_{i} = (o_{i1}, \cdots, o_{iC}) \in \mathbbm{R}^{C}$ for all $i \in \{1, \cdots, L\}$ where $o_{ij} = \log p(y_i = j | \hat{y}_{<i}, x, \theta)$. Our intuition behind the beam search-based logit acquisition is that \textit{considering the logits obtained from beam search for adaptation naturally aligned with how ASR models decode sentences}, \ie, our approach can tailor the model to adapt toward the actual yet accurate sentences generated by the ASR models.

\subsection{Generalized Entropy Minimization}
\label{subsec:gem}
While entropy minimization achieves decent performance on domain adaptation tasks~\cite{wang2020tent, fleuret2021test, zhang2021memo, khurana2021unsupervised} by reducing the uncertainty of predictions, and extracting domain-invariant features on the target domain, we can further improve this objective by adopting its generalized version, R\'enyi entropy, and searching for more effective hyperparameter. For a discrete random variable $X$, which takes values in $\{1, \cdots, C \}$, R\'enyi entropy $\text{H}_{\alpha}(X)$ of order $\alpha$ with $\alpha \in (0, 1) \cup (1, \infty)$ is defined as $\text{H}_{\alpha}(X) = \frac{1}{1 - \alpha}{\log \big( \sum_{j=1}^{C} \mathbb{P}(X = j)^{\alpha} \big)}$.
When $\alpha \rightarrow 1$ and $\alpha \rightarrow \infty$, $\text{H}_{\alpha}(X)$ becomes Shannon entropy and cross-entropy with a pseudo-label $\argmax_{j} \mathbb{P}(X = j)$, respectively. For a single-utterance TTA setting, we hypothesize that there exists an optimal $\alpha \in (1, \infty)$ and define the generalized entropy loss as follows:
\begin{equation}
\mathcal{L}_{\text{GEM}} = \frac{1}{L}\sum_{i=1}^{L} \frac{1}{1 - \alpha} \log \Big( \sum_{j=1}^{C} p_{ij}^{\alpha} \Big),
\end{equation}
where $p_{ij} = \frac{\exp(o_{ij} / T)}{{\sum_{j' = 1}^{C}{\exp(o_{i j'} / T)}}}$ and $T$ is a temperature hyperparameter for preventing \textit{vanishing gradient}. As the blank token dominates all timesteps, we ignore timesteps with the highest probability of blank token among all classes to alleviate the class imbalance problem as in~\cite{suta}. 

\subsection{Negative Sampling}
\label{subsec:ns}
We further exploit negative sampling loss, originally adopted for semi-supervised learning (SSL) in~\cite{chen2020negative}. Negative sampling loss penalizes the probabilities of low-confident classes, and Chen et al.~\cite{chen2020negative} have shown that adding it can further boost the performance of existing SSL algorithms. It can be derived from the standard cross-entropy loss as follows. Given $L$ labeled samples ${\{(x_i,  y_i) \}}_{i=1}^{L}$, the standard cross-entropy loss is defined as $\mathcal{L}_{\text{CE}} = -\frac{1}{L}\sum_{i = 1}^{L}\sum_{j=1}^{C} \mathbbm{1}_{[j = y_i]} \log p_{ij}$. Note that $\sum_{j=1}^{C} \mathbbm{1}_{[j = y_i]} \log p_{ij} = \log ( 1 - \sum_{j \ne \text{$y_i$}} p_{ij} )$. Since we do not know the ground truth label $y_i$ for each $x_i$ in the unlabeled target domain, we approximate $\mathcal{L}_{\text{CE}}$ with the negative sampling loss $\mathcal{L}_{\text{NS}}$ defined as follows:
\begin{equation}
\label{eq:negative_sampling}
\mathcal{L}_{\text{NS}} = - \frac{1}{L}\sum_{i=1}^{L} \log \Big(1 -  \sum_{j=1}^{C} \mathbbm{1}_{[p'_{ij} < \tau]} p_{ij} \Big),
\end{equation}
where $p_{ij} = \frac{\exp(o_{ij} / T)}{{\sum_{j' = 1}^{C}{\exp(o_{i j'} / T)}}}$, $p'_{ij} = \frac{\exp(o_{ij})}{{\sum_{j' = 1}^{C}{\exp(o_{i j'})}}}$ with a temperature hyperparameter $T$ for avoiding vanishing gradient, and $\mathbbm{1}$ is an indicator function. $j$-th class of $x_i$ is considered as a negative class when the probability $p'_{ij}$ is less than a threshold $\tau$. Without modification, Equation \eqref{eq:negative_sampling} can be interpreted in a single-utterance TTA setting as penalizing probabilities of negative classes at every timestep for a sequential output of length $L$.

\subsection{Overall Framework}
\label{subsec:overall_framework}
Our final unsupervised objective is the weighted sum of the generalized entropy loss and the negative sampling loss as follows:
\begin{equation}
\label{eq:overall}
\mathcal{L} = \mathcal{L}_{\text{GEM}} + \lambda_{\text{NS}} \mathcal{L}_{\text{NS}},
\end{equation}
where $\lambda_{\text{NS}}$ is negative sampling weight for balancing two losses. For each utterance, we adapt the model for $N$ iterations in an \textit{episodic manner} where we newly reset the model to the pre-trained one to preserve the knowledge from the source domain.

\begin{table*}[!t]
\scriptsize
\centering
\caption{Comparison of TTA performance, measured as word error rate (\%), for the CTC-based model across 12 datasets with various domain shifts. The results are obtained using beam search decoding with an external language model for inference.}
\label{table:main_beam}
\vspace{-0.1in}
\begin{tabular}{lccccccccccccc}
\Xhline{3\arrayrulewidth}
    \textbf{Dataset} & \textbf{CH} & \textbf{TD} & \textbf{CV} & \textbf{VA} & \textbf{AC} & \textbf{AA} & \textbf{BA} & \textbf{CM} & \textbf{MU} & \textbf{NB} & \textbf{SD} & \textbf{TP} & \textbf{Avg.} \\ \hline
    Unadapted & 29.5 & 12.2 & 36.9 & 13.0 & 26.1 & 38.6 & 58.9 & 48.9 & 49.0 & \textbf{91.6} & 17.4 & 23.7 & 37.2 \\ \hline
    SUTA & \textbf{24.1} & \textbf{11.6} & 31.5 & 11.4 & 16.8 & 30.3 & 53.2 & 38.1 & 38.6 & 107.9 & 14.1 & \textbf{16.9} & 32.9 \\ \hline
    SGEM & \textbf{24.1} & 11.7 & \textbf{31.1} & \textbf{11.1} & \textbf{16.5} & \textbf{29.8} & \textbf{51.6} & \textbf{37.7} & \textbf{37.7} & 106.7 & \textbf{14.0} & \textbf{16.9} & \textbf{32.4} \\ 
\Xhline{3\arrayrulewidth}
\vspace{-0.3in}
\end{tabular}
\end{table*}

\section{Experiments}
\label{sec:experiments}
\subsection{Experimental Setup}
\noindent\textbf{Source ASR Models}\quad To verify the efficacy of SGEM, we evaluate it on three mainstream ASR architectures: the CTC-based model~\cite{graves2006connectionist}, Conformer~\cite{gulati2020conformer}, and Transducer~\cite{graves2012sequence}. More specifically, for the CTC-based model, we use wav2vec 2.0~\cite{baevski2020wav2vec} \footnote{\scriptsize{\url{{https://huggingface.co/facebook/wav2vec2-base-960h}}}} trained on the LibriSpeech dataset~\cite{panayotov2015librispeech}. For Conformer, we exploit Conformer-CTC~\cite{gulati2020conformer} \footnote{\scriptsize{\url{https://catalog.ngc.nvidia.com/orgs/nvidia/teams/nemo/models/stt_en_conformer_ctc_small_ls}}} trained on the LibriSpeech dataset. For Transducer, we adopt Conformer-Transducer~\cite{burchi2021efficient} \footnote{\scriptsize{\url{https://catalog.ngc.nvidia.com/orgs/nvidia/teams/nemo/models/stt_en_conformer_transducer_small}}} trained on a composite NeMO ASRSET dataset, including the LibriSpeech dataset. We utilize the external 4-gram language model \footnote{\scriptsize{\url{https://huggingface.co/patrickvonplaten/wav2vec2-base-100h-with-lm}}} for the CTC-based model and Conformer.
\\
\\
\noindent\textbf{Datasets}\quad We assess the performance of SGEM on multiple datasets under various domain shift settings. To test SGEM under unseen speakers/words, we use the test set of four datasets: CHiME-3 (CH)~\cite{barker2015third}, TED-LIUM 2 (TD)~\cite{rousseau2014enhancing}, Common Voice (CV)~\cite{ardila2019common}, and Valentini (VA)~\cite{valentini2017noisy}. In addition, we validate SGEM under accident background noise by injecting the following eight types of noises to each utterance of in-domain LibriSpeech test-other dataset~\cite{panayotov2015librispeech}: air conditioner (AC), airport announcement (AA), babble (BA), copy machine (CM), munching (MU), neighbors (NB), shutting door (SD), and typing (TP) with $\text{SNR} = 10$dB. For each type of noise, we randomly select one noise sample from the MS-SNSD noise test set~\cite{reddy2019scalable}. We also evaluate SGEM on L2-Arctic~\cite{zhao2018l2arctic}, non-native English speech corpora, to verify SGEM under extreme pronunciation/accent shifts. Specifically, we randomly select one speaker for each first language.
\\
\\
\noindent\textbf{Implementation Details}\quad Since the TTA setting has no validation set, we optimize hyperparameters on the CH dataset for each model and apply them to the other datasets. The best settings are as follows. For all models, we use AdamW optimizer~\cite{loshchilov2017decoupled} and cosine annealing learning rate scheduler with $\eta_i$ and $\eta_f$ for initial and final learning rates, respectively, and set $(N, T, \tau) = (10, 2.5, 0.4/C)$ with vocabulary size $C$. We only train feature extractors for the CTC-based model and encoders for the others. Furthermore, we set $(\eta_i, \eta_f, B, \lambda_{\text{LM}}, \alpha, \lambda_{\text{NS}})=(4 \blank 10^{-5}, 2 \blank 10^{-5}, 5, 0.3, 1.5, 1)$ for the CTC-based model, $(4 \blank 10^{-5}, 2 \blank 10^{-5}, 5, 0.3, 1.25, 2)$ for Conformer, and $(4 \blank 10^{-6}, 2 \blank 10^{-6}, 3, 0, 1.25, 0.5)$ for Transducer. All experiments are conducted on Nvidia TITAN Xp and GeForce RTX 3090. Adaptation takes about 0.771 seconds for a 1-second utterance averaged over three models.

\subsection{Main Results}
We compare the TTA performance of three mainstream ASR models, including the CTC-based model, Conformer, and Transducer, across 12 datasets with various domain shifts. Table \ref{table:main} presents the word error rate (WER) of ASR model outputs generated by the greedy search decoding method, following the evaluation protocol used in the previous study~\cite{suta}. Additionally, Table \ref{table:main_beam} showcases the TTA performance for the CTC-based models using beam search decoding with external LM. For both decoding methods, the ASR models with SGEM consistently enhance the recognition accuracy of target utterances with an average word error rate reduction of 15.6\%, except for two cases on NB, where the performance without adaptation is the best when using beam search decoding. Furthermore, SGEM outperforms SUTA in terms of average WER across all 12 datasets for each of the three model architectures (\textbf{CTC-based model}: (greedy) $34.1\% \rightarrow \textbf{33.4\%}$, (beam search) $32.9\% \rightarrow \textbf{32.4\%}$ / \textbf{Conformer}: $39.3\% \rightarrow \textbf{38.4\%}$ / \textbf{Transducer}: $20.8\% \rightarrow \textbf{20.6\%}$). This indicates the superiority of our unsupervised objectives as well as the logit acquisition method for adapting sequential language outputs regardless of the decoding strategy.

\subsection{Non-Native English Speech Corpora} \label{subsec:non-native}
To show the usability of SGEM at various domain shifts, we further analyze SGEM on six different non-native English speech corpora. The result is summarized in Table \ref{table:non_native}. As shown in Table \ref{table:non_native}, SGEM achieves the best results for all corpora, outperforming the baseline. This implies the adaptability of SGEM under extreme pronunciation/accent shifts, demonstrating its versatility in practical situations with severe speaker shifts, such as globally used online ASR systems.

\begin{table}[!h]
\scriptsize
\centering
\vspace{-0.1in}
\caption{Comparison of TTA performance, measured as word error rate (\%), for the CTC-based model on six non-native English speech corpora.}
\label{table:non_native}
\vspace{-0.1in}
\begin{tabular}{lccc}
\Xhline{3\arrayrulewidth}
    \textbf{Setting} & \textbf{Unadapted} & \textbf{SUTA} & \textbf{SGEM} \\ \hline
    Arabic & 32.5 & 27.1 & \textbf{26.5} \\
    Mandarin & 28.5 & 23.3 & \textbf{23.1} \\
    Hindi & 15.7 & 12.5 & \textbf{12.3} \\
    Korean & 23.3 & 19.7 & \textbf{19.5} \\
    Spanish & 35.7 & 29.8 & \textbf{29.3} \\
    Vietnamese & 18.5 & 15.7 & \textbf{15.4} \\ \hline
    \textbf{Average} & 25.7 & 21.4 &  \textbf{21.0} \\
\Xhline{3\arrayrulewidth}
\vspace{-0.25in}
\end{tabular}
\end{table}

\subsection{Data Deficient Condition}
\begin{figure}[t]
    \centering
    \includegraphics[width=0.25\textwidth]{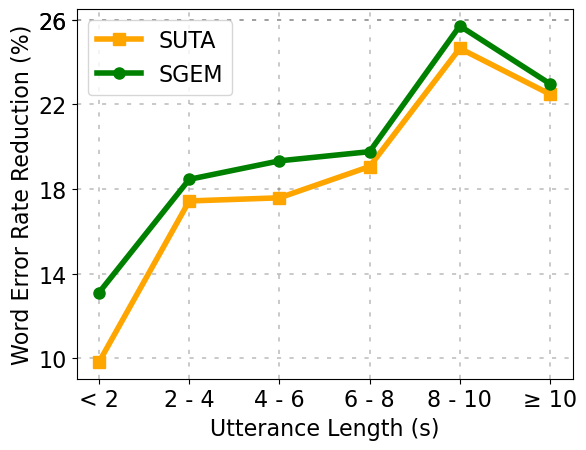}
    \vspace{-0.15in}
    \caption{Comparison of TTA performance, measured as relative word error rate reduction (\%), on different utterance lengths of CH dataset for the CTC-based model.}
    \label{fig:data_deficient_condition}
    \vspace{-0.25in}
\end{figure}

It is commonly known that TTA methods fail under the data deficient condition where the number of test instances is limited~\cite{niu2023towards, gao2022back}. This still holds in the single-utterance TTA setting for ASR models, where the length of utterance is short, so the number of output tokens is insufficient. To validate SGEM on this harsh condition, we split the CH dataset according to utterance length and evaluate SGEM with the CTC-based model on each split. As shown in Figure \ref{fig:data_deficient_condition}, SGEM performs best in every length interval. In addition, it is worth noting that SGEM significantly outperforms the baseline for extremely short utterances of less than 2 seconds, showing the superiority of our method in real-world scenarios where short utterances are prevalent and negligible latency is required.

\subsection{Ablation Study}
To validate the core components of SGEM: beam search-based logit acquisition (\textbf{BS}, Section \ref{subsec:logit_acquisition}), generalized entropy minimization (\textbf{GEM}, Section \ref{subsec:gem}), and negative sampling (\textbf{NS}, Section \ref{subsec:ns}), we conduct an ablation study for three mainstream ASR models on CH dataset. As shown in Table \ref{table:ablation}, both generalized entropy minimization and negative sampling alone achieve remarkable performance gains for every model, indicating the efficacy of each component. Meanwhile, substituting greedy search for beam search even with small beam width (for all models) and without external LM (for Transducer) consistently boosts the performance in all cases, showing the effectiveness of beam search-based logit acquisition and implying that additional performance improvement can be expected using larger beam sizes or language model if resources are allowed.

\begin{table}[h]
\scriptsize
\centering
\caption{Ablation study of SGEM.}
\label{table:ablation}
\vspace{-0.1in}
\setlength{\tabcolsep}{5pt}
\begin{tabular}{ccc|ccc}
\Xhline{3\arrayrulewidth}
    \textbf{BS} & \textbf{GEM}  & \textbf{NS} & \textbf{CTC} & \textbf{Conformer} & \textbf{Transducer} \\ \hline
    \lightgrayx & \lightgrayx & \lightgrayx & 31.2 & 28.7 & 11.8 \\ \hdashline
    \lightgrayx & \cmark & \lightgrayx & 24.9 & 24.7 & 10.0 \\
    \lightgrayx & \lightgrayx & \cmark & 25.2 & 25.0 & 10.1 \\
    \lightgrayx & \cmark & \cmark & 24.8 & 24.7 & 10.0 \\
    \cmark & \lightgrayx & \cmark & 24.8 & 24.7 & 10.1 \\ \hline
    \cmark & \cmark & \cmark & \textbf{24.7} & \textbf{24.5} & \textbf{9.9} \\
\Xhline{3\arrayrulewidth}
\end{tabular}
\vspace{-0.2in}
\end{table}

\section{Conclusion}
We have suggested SGEM, an effective single-utterance TTA framework for general ASR models. SGEM exploits beam search-based logit acquisition and utilizes generalized entropy minimization and negative sampling objectives to adapt the model at a sequential level. SGEM has achieved state-of-the-art performance for the three mainstream ASR models across 12 datasets with various domain shifts, including utterances of unseen speakers/words during training and utterances with severe background noise. We have also verified SGEM under harsh conditions, such as non-native English utterances with severe pronunciation/accent shifts and the data deficient condition, and the efficacy of each component of SGEM through an ablation study. SGEM sheds light on the careful design of speech-specific components when devising test-time adaptation methods for ASR models.

\section{Acknowledgement}
This work was supported by National Research Foundation of Korea (NRF) grants (No.2018R1A5A1059921, No.2019R1C1C1009192) and Institute of Information \& communications Technology Planning \& Evaluation (IITP) grant funded by the Korea government (MSIT) (No.2019-0-00075, Artificial Intelligence Graduate School Program(KAIST)).

\bibliographystyle{IEEEtran}
\bibliography{mybib}

\end{document}